\documentstyle[seceq,epsf]{ptptex}

\title{
Evolution of Primordial Protostellar Clouds \\
---{\it Quasi-Static Analysis}---}

\author{
Kazuyuki {\sc Omukai},
Ryoichi {\sc Nishi},
Hideya {\sc Uehara}
and Hajime {\sc Susa}$^*$
}

\inst{
Department of Physics, Kyoto University, Kyoto 606-8502, Japan 
\\
$^*$ Center for Computational Physics, University of Tsukuba\\
           Tsukuba 305-8577, Japan
}

\recdate{
\today
}

\abst{
  The contraction processes of metal-free molecular clouds of starlike
  mass (or cloud cores) are investigated.   
  We calculate radiative transfer of the H$_2$ lines 
  and examine quasi-static contraction with radiative cooling.  
  Comparing two time-scales, the free-fall time 
  $t_{\rm{ff}}$ and the time-scale of
  quasi-static contraction $t_{\rm{qsc}}$ ($\sim t_{\rm{cool}}$, the cooling
  time) of these cores, we find that
  the ratio of the two time-scales $t_{\rm{ff}}/t_{\rm{qsc}}$,
  i.e., the efficiency of cooling, becomes larger 
  with contraction even under the
  existence of cold and opaque envelopes. In particular, for 
  fragments of primordial filamentary clouds, for which $t_{\rm{ff}} \sim t_{\rm{qsc}}$
  at the fragmentation epoch, they collapse dynamically in the free-fall time-scale. This
  efficiency of cooling is unique to line cooling.
}

\begin{document}

\maketitle

\section{Introduction}
One of the most conspicuous features of galaxies is the existence of
stars.
To discuss the formation and evolution of galaxies, then, 
it is indispensable to investigate star formation processes in
protogalactic clouds.
In addition, stars play crucial roles in galactic activities, 
for example, ultraviolet radiation and supernova explosions, which 
are important in some cosmological contexts.
Recently, considering the hierarchical clustering scenario, 
several authors stressed the importance of stars in pregalctic objects  
for the reionization of the universe (e.g., Ref.~\citen{rf:1}).  
However, the formation of first stars is one of the most poorly
understood processes involved in galaxy formation.
For this reason, theoretical predictions on the reionization by first
stars suffer great uncertainty. 
Therefore it is definitely well-timed to investigate the formation process
of first stars. 

The evolution of primordial clouds from which first stars form has been
investigated with a one-zone approximation by many authors.\cite{rf:2}
However, the mass scale of the clouds which are able to collapse is much
larger than the typical stellar mass.  
Then, we must discuss the fragmentation processes of
protogalactic clouds and the evolution of these fragments into stars to 
understand the star formation. 

With respect to the fragmentation process, several authors have envisaged that
protogalactic clouds first collapse disk-like, and then fragment into 
filaments (e.g., Ref.~\citen{rf:3}). 
Uehara et al.\cite{rf:4} investigated the collapse of
filamentary primordial clouds with a one-zone approximation and concluded
that the minimum mass of the fragments becomes $\sim 1M_{\odot}$, that is,
essentially the Chandrasekhar mass.  
Their calculation also showed that, for low mass fragments, almost all of
hydrogen atoms are converted into molecular form before the
fragmentation occurs.  

On the other hand, Stahler et al.\cite{rf:5}
studied the main accretion phase of a primordial protostar assuming a
very low mass hydrostatic core (i.e., a stellar core, see
Ref.~\citen{rf:6}) initially, and stationary accretion of the envelope
onto it.  
They adopted a higher mass accretion rate $(4.41 \times 10^{-3} M_{\odot}
~{\rm yr}^{-1})$ than the present-day counterpart ($10^{-5} M_{\odot}
~{\rm yr}^{-1}$, e.g., Ref.~\citen{rf:7}). 
This higher mass accretion rate is caused by higher temperature 
$\sim 10^3$ K of primordial gas clouds. 
However, the question of how a fragment of star-like mass reaches this
phase is not considered and still remains unclear.  
In this paper, we study the evolution of stellar mass primordial clouds
shortly after the fragmentation.
  
How primordial clouds of stellar mass, which are fragments of
filamentary clouds, evolve after the fragmentation depends on their
efficiency of cooling, i.e., the ratio of the free-fall time-scale
$t_{\rm{ff}}$ to the cooling time-scale $t_{\rm{cool}}$ (e.g.,
Ref.~\citen{rf:8}).   
If $t_{\rm{ff}}<t_{\rm{cool}}$ (i.e., cooling is not effective), while
the cloud cools, it has enough time to adjust itself to a new
hydrostatic configuration. 
Then it contracts, maintaining a nearly hydrostatic equilibrium (i.e., 
Kelvin-Helmholtz contraction). 
On the other hand, in the case $t_{\rm{ff}}>t_{\rm{cool}}$, the cloud
collapses dynamically in its free-fall time-scale after substantial
cooling, and no hydrostatic equilibrium is established. 
In the latter case, the cloud may fragment into smaller pieces again.

At the time of the fragmentation of filamentary clouds, the two time-scales of
fragments, $t_{\rm{cool}}$ and $t_{\rm{ff}}$, are of the same order of
magnitude,\cite{rf:4} since a filamentary cloud fragments
when the time-scale of contraction $t_{\rm{dyn}} \equiv \rho /(d \rho /dt)$ 
becomes nearly equal to the time-scale of fragmentation 
$t_{\rm{frag}}\equiv 2.1/\sqrt{2\pi G \rho}$,\cite{rf:9} where
$\rho$ is the density of the cloud.   
At the epoch of fragmentation, the contraction proceeds
isothermally with regard to time. 
Then $t_{\rm{dyn}} \simeq 
(\gamma_{\rm{ad}}-1)t_{\rm{cool}}$, where the adiabatic exponent
$\gamma_{\rm{ad}}$ is 5/3 for monatomic molecules and 7/5 for diatomic
molecules.   
Since the fragmentation time-scale $t_{\rm{frag}}$ of a filamentary
cloud is identical (up to numerical factors) to the free-fall time-scale 
$t_{\rm{ff}}$ of a spherical fragment, $t_{\rm{cool}} \sim t_{\rm{ff}}$ for
the fragment at the fragmentation epoch. 
Therefore, the nature of the evolution depends on whether the ratio
$t_{\rm{cool}}/t_{\rm{ff}}$ [more precisely, $t_{\rm{qsc}}/t_{\rm{ff}}$,
where $t_{\rm{qsc}}$ is the time-scale of quasi-static contraction (see
\S 3 below)] becomes larger or smaller as a fragment contracts.  

To estimate the cooling time, we must investigate the cooling processes in 
the primordial clouds. 
The principal difference of a primordial star forming region from its
present-day counterpart is its lack of heavy elements and
consequently dust grains, which are efficient coolants, keeping
present-day molecular clouds almost isothermal $(\sim 10~{\rm K})$ over
many orders of magnitude in density (e.g., Ref.~\citen{rf:10}).
Therefore, when we investigate primordial star formation, it is 
no longer possible to impose isothermality. 
We must study cooling/heating processes and solve the energy equation
to determine the temperature.
At low temperature $(T<10^{4}~{\rm K})$, the Ly$\alpha$ line of the
neutral hydrogen is hardly excited, and there is no atomic cooling process. 
If there are H$_{2}$ molecules, however, these clouds can cool by
H$_{2}$ rotational/vibrational transition lines.  
According to Uehara et al.,\cite{rf:4} at the time of fragmentation,
filamentary clouds are optically thick for line center frequency of some
lines.
This should also hold for fragments of them.  
The surface temperature of the clouds is low, in comparison with the
central or averaged temperature. 
Consequently, the efficiency of cooling by optically thick lines may decrease. 
For example, the continuum energy flux decreases as $T_{\rm{eff}}^{4}$,
where $T_{\rm{eff}}$ is the effective (photospheric) temperature of a
cloud. 
In an optically thick case, the photosphere is located near the surface, and
the low surface temperature causes a severe decrease in the cooling rate. 
However, in the case that cooling is dominated by optically thick lines,
optical depth is not only different from one line to another line, but it
also varies with frequency even within one line.  
In fact, some lines are optically thick at their line center frequencies,
but clouds are totally transparent to intervening 
frequencies between lines. 
Therefore the notion of a photosphere cannot be easily implimented. 
Final processed line profiles depend on the structure of the cloud and
cannot be known before we calculate radiative transfer resolving
frequencies within a single line. 
Moreover, the number of effective lines changes as the cloud contracts. 
For these reasons, to understand the evolution of the fragments,
it is necessary to treat the H$_{2}$ line transfer problem in detail. 

The outline of this paper is as follows. 
In \S 2, we describe our scheme of the calculation of the luminosity
and the cooling rate via the H$_2$ emission lines for spherical clouds
and give some results. 
In \S 3, we compare the time-scales $t_{\rm{ff}}$ and $t_{\rm{cool}}$
for general initial conditions and discuss the 
evolution of fragments of filamentary clouds. 
Some interpretations of the results are also 
presented in \S 4. 
Finally, we summarize our work and discuss its implications with respect 
to the primordial star formation in \S 5. 

\section{Radiative processes in primordial molecular cloud cores}
We consider spherically symmetric gas clouds of starlike mass ($\sim 1
M_{\odot}$) (`molecular cores') composed only of H$_2$ molecules, because
the hydrogen atoms are converted into molecular form before the
fragmentation for these low mass fragments.\cite{rf:4}   
We can neglect helium, since it is thermally inert at these low
temperatures.  
As mentioned above, these clouds lose their thermal energy by radiative
cooling via the H$_2$ rotovibrational lines, and presumably contract
into protostars.  
 
In this section, we describe our calculation scheme for the luminosity and
cooling rate of a spherically symmetric cloud. 
The specific intensity $I_{\nu}~({\rm ergs}~{\rm sec^{-1}} {\rm
  cm^{-2}}{\rm sr^{-1}}{\rm Hz^{-1}})$ along a ray is calculated by
solving the radiative transfer equation (e.g.,~Ref.~\citen{rf:11}), 
\begin{equation}
\frac{dI_{\nu}}{ds}=-\alpha_{\nu}I_{\nu}+j_{\nu} .
\label{eq:tr}
\end{equation}
Here $s$ is the displacement along the ray, and $\alpha_{\nu}$ and  
$j_{\nu}$ are the absorption and emission coefficients, respectively. 
These coefficients can be written by using Einstein $A-$ and
$B-$coefficients:
\begin{eqnarray}
j_{\nu}&=&\frac{h \nu}{4 \pi} n_{2} A_{21} \phi (\nu),\\
\alpha_{\nu}&=&\frac{h \nu}{4 \pi}\phi (\nu) (n_{1}B_{12}-n_{2}B_{21}).
\end{eqnarray}
Here $n_{1}$ and $n_{2}$ are the number densities of the molecules in
the lower and upper level of the transition, respectively, and
$\phi(\nu)$ is the line profile function.  
In our calculation, the line center optical depth of the lines across a
cloud $\tau_{\rm{c}}$ is no more than several hundreds. 
Since Lorentz wings of a line become more important than its Doppler
core only when $\tau_{\rm{c}}$ is as large as $10^{3}$, we only consider
line broadening owing to the thermal Doppler effect, i.e.,  
\begin{equation}
\phi(\nu)=\frac{1}{\Delta {\nu}_{\rm{D}} \sqrt{\pi}} e^{-(\nu-\nu_{0})^{2}
  /(\Delta {\nu}_{\rm{D}})^{2} }.
\end{equation}
Here ${\nu}_{0}$ is the line center frequency, and the Doppler width
$\Delta {\nu}_{\rm{D}}$ is defined by
\begin{equation}
\Delta {\nu}_{\rm{D}}=\frac{{\nu}_{0}}{c}\sqrt{\frac{2kT}{\mu m_{\rm H}}}, 
\end{equation}
where $T$ is the temperature at each radius in the cloud, $m_{\rm
  H}=1.67 \times 10^{-24}~{\rm g}$ is the mass of a hydrogen atom, and
  $\mu=2$ is the molecular weight of the molecular hydrogen.  
We include the first three vibrational states, with rotational levels up 
to $j=20$ in each state, following Palla et al.\cite{rf:2}  
We take $A$-coefficients of molecular hydrogen from Turner et
  al.\cite{rf:12}  
$B$-coefficients can be obtained from $A$-coefficients by the following
  Einstein relations, 
\begin{eqnarray}
g_{1}B_{12}&=&g_{2}B_{21},\\
A_{21}&=&\frac{2h {\nu_0}^{3}}{c^{2}} B_{21},
\end{eqnarray}
where $g_{i}$ is the statistical weight of the level $i$. 
The critical number density $n_{\rm{cr}}$, where radiative and
collisional deexicitation rates become equal for molecular hydrogen, is
$ \sim 10^{4}~{\rm cm^{-3}}$.  
For a typical number density of fragments $(\sim 10^{10}~{\rm cm^{-3}}
\gg n_{\rm{cr}})$, almost all excited molecules would be deexcited by
collision with other molecules. 
Therefore, collisional local thermodynamic equilibrium (LTE) is
established, and scattering of photons can be neglected. 
These properties make the transfer problem of H$_2$ lines fairly tractable. 

Figure 1 displays the geometry of grids of the radiative transfer calculation 
in the case of spherical symmetry (Ref.~\citen{rf:13}; see also
Ref.~\citen{rf:14}).   
To know the luminosity at the $i$-th radial grid $r_{i}$, we solve the
transfer equation (\ref{eq:tr}) with frequency $\nu$ along from
the 0-th to $(i-1)$-th rays, and we obtain the specific intensity at $r_{i}$ in
the direction $\theta_{ij}$, $I_{\nu}(r_{i},\theta_{ij}) ~(-i+1
\le j \le i-1)$. 
This procedure will be repeated for each frequency mesh. 
We consider the frequency range for each line within
four times the Doppler width corresponding to the central 
temperature from the line center. 
Each line is usually separated into 20 frequency meshes (in some cases,
e.g., Fig. 7, 80 frequency meshes), and the number of radial meshes
$n=50$ in our runs. 
Integrating over $\theta$ and $\nu$, we obtain the
monochromatic energy flux $F_{\nu}$ and the luminosity $L$: 
\begin{eqnarray}
F_{\nu}(r)&=&2 \pi \int_{0}^{\pi} I_{\nu}(r,\theta){\rm cos} \theta
d\theta,\\
L(r)&=&4\pi r^{2} \int_{0}^{\infty} F_{\nu}(r) d \nu .
\end{eqnarray}

As for cooling rate per unit mass $\Lambda(m)$, we differentiate the
luminosity $L$ with respect to the Lagrangian mass coordinate $m$,
\begin{eqnarray}
\Lambda(m)&=&\frac{\partial L}{\partial m},\\
m(r)&=&\int_{0}^{r} 4\pi r'^{2} \rho dr' .
\end{eqnarray}  

Figure 2 displays the contribution from each line to the luminosity at the
surface of a typical cloud. 
The mass of the cloud is $1 M_{\odot}$, the number density at half mass
radius is $10^{11}~{\rm cm}^{-3}$, and the density/temperature
distribution is represented by the Emden function of polytropic index
2.5. 
In this spectrum, there is a gap around the wavelength of $4 \mu$m, and
lines can be classified into longer and shorter wavelength groups. 
The shorter and longer wavelength lines correspond to transitions with
and without changes of vibrational levels.     
\section{Evolution of fragments} 
As initial conditions, we assume polytropic gas spheres in hydrostatic
equilibrium; namely the density/temperature distribution is represented
by the Emden function of polytropic index $N$.
Clouds are cut off at the radius $r_{s}$, where the density
falls off by a factor of $10^{-3}$ from the central value. 
Therefore, parameters characterizing initial conditions are the effective
polytropic index $N$, the total mass of the cloud $M$, and the number
density $n_{h}$ at the half mass radius. 
For a given initial configuration, we can obtain the specific entropy
distribution $s(m,t=0)$.   

We then bring forward the cloud one time step $\Delta t$ {\em
  quasi-statically} in the following way.    
We calculate the cooling rate $\Lambda(m)$ as described in \S 2, 
and advance the specific entropy distribution $s(m,t)$
by one time step $\Delta t$ using the heat equation  
\begin{equation}
\frac{\partial s}{\partial t}=-\frac{1}{T} \Lambda(m).
\end{equation}
With the entropy distribution at the next time step $s(m,t+\Delta
t)=s(m,t)-(\Lambda(m)/T)\Delta t$, we find the new equilibrium
configuration using the equations of hydrostatic equilibrium, 
\begin{eqnarray}
\frac{\partial r}{\partial m}&=&\frac{1}{4 \pi r^{2} \rho },\\
\frac{\partial p}{\partial m}&=&-\frac{G m}{4 \pi r^{4}},
\end{eqnarray}
and the equation of state,
\begin{equation}
p \propto {\rm exp}[s/c_{\rm v}] {\rho}^{\gamma_{\rm{ad}}}.
\end{equation}
Here $\gamma_{\rm{ad}}=7/5$ is the adiabatic coefficient and 
$c_{\rm v}=\frac{1}{\gamma_{\rm{ad}}-1}\frac{k}{\mu m_{\rm H}}$ is
specific heat of hydrogen molecule at constant volume. 
We impose a constant external pressure at the outer boundary. 
For dense and cold clouds, like the fragments under consideration, radiation
pressure can be neglected, and the pressure can be assumed to be given
by the gas pressure only. 

For given states of clouds, we compare two time-scales, i.e., the
free-fall time
\begin{equation}
t_{\rm{ff}}(m)\equiv \sqrt{\frac{3\pi}{32 G \bar{\rho}(m)}},
\end{equation}
and the time-scale of quasi-static contraction,  
\begin{equation}
t_{\rm{qsc}}(m)\equiv \rho /\left(\frac{\partial \rho}{\partial t}
\right)_{m,{\rm quasi \hbox{-}static}}.
\end{equation}
Here, $\bar{\rho}(m)$ is the mean density within the mass coordinate
$m$, and
each quantity is evaluated at the coordinate $m$. From the virial theorem,
$t_{\rm{qsc}}$ is of the same order of magnitude of $t_{\rm{cool}}$.

Figure 3 shows the contours  of 
the ratio $t_{\rm{ff}}/t_{\rm{qsc}}$ at the center ($m=0$) of clouds with
(a) $N=2.5$ (adiabatic stratification) and (b) $N=8$. 
Regions around the curves $t_{\rm{ff}}/t_{\rm{qsc}}=1$ are expected to be 
initial states of fragments. 
Although Fig. 3 displays the values at the
center, this ratio does not change significantly within a cloud, except
near the surface. 
The upper-right regions correspond to higher central temperature $>
2000~{\rm K}$, where the molecular hydrogen begins to dissociate. 
From the two panels of Fig. 3, we can see that the ratio
$t_{\rm{ff}}/t_{\rm{qsc}}$ depends only slightly on the effective polytropic
indices $N$, rather than on masses and densities. 
Note that for clouds of the same mass, the ratio
$t_{\rm{ff}}/t_{\rm{qsc}}$ is greater for denser cloud. 
This implies that as a cloud contracts, the ratio $t_{\rm{ff}}/t_{\rm{qsc}}$
becomes larger. 
In particular, for fragments of filamentary gas clouds, for which
$t_{\rm{ff}} \sim t_{\rm{qsc}}$ initially, $t_{\rm{qsc}}$ becomes
shorter than $t_{\rm{ff}}$ as they contract; i.e., such clouds collapse
dynamically. 
Note that when $t_{\rm{qsc}}<t_{\rm{ff}}$, clouds do not contract 
quasi-statically, and $t_{\rm{qsc}}$ does not possess the meaning of the
collapse time-scale. 
In this case, $t_{\rm{qsc}}$ merely measures the time-scale of cooling.  

The discussion above is for fixed $N$.
However, these clouds change their entropy distribution as they contract. 
Therefore it is not enough to discuss their evolution only with fixed $N$. 
We need to calculate the evolution of clouds contracting quasi-statically
and to show that they in fact begin to collapse dynamically. 
We choose initial conditions under which clouds contract
quasi-statically and pursue their evolution. 
Their initial parameters are (a)~$M=0.5M_{\odot},~n_{h}=10^{9}~{\rm
  cm^{-3}},~N=2.5$ (the asterisk $\ast$ in Fig.3(a)), and
(b)~$M=0.2M_{\odot},~n_{h}=10^{9}~{\rm cm^{-3}},~N=20$. For the
parameter representing the structure of clouds, we employ the effective
exponents for structure,
\begin{equation}
\gamma_{\rm str}(m) \equiv \frac{(\partial {\rm ln} p/ \partial
  m)_{t}}{(\partial {\rm ln} \rho / \partial m)_{t}}.
\end{equation}

Figure 4 displays changes of the $\gamma_{\rm str}(m)$ as the clouds
 contract.
We can see that values of $\gamma_{\rm str}(m)$ become nearly flat
 and approach $1.12-1.13$ when the central temperature reaches $\sim
 2000~{\rm K}$, which is the dissociation temperature of  H$_2$.
 These values of $\gamma_{\rm str}$ correspond to $N \simeq 8$, regardless
 of the initial $N$ values. However, for the fragments, 
 the clouds start to collapse dynamically before these states are 
 reached. 
 In such a case, this limiting value of
 $N$ has no significant meaning. But the very existence of the
 limiting value
 supports our discussion above with fixed $N$. In fact,  
 from Fig. 5, where the change of time-scales and their ratio are shown
 for the case (a) of Fig. 4,
we can see that the cloud indeed collapses dynamically after it contracts 
quasi-statically to some extent.  

Although $\gamma_{\rm str}$ becomes $\simeq 1.13$, the 
effective exponent for temporal evolution of a fixed mass
element, 
\begin{equation}
\gamma_{\rm ev}(m) \equiv \frac{(\partial {\rm ln} p/ \partial
  t)_{m}}{(\partial {\rm ln} \rho / \partial t)_{m}},
\end{equation}
has different value. 
The distributions of $\gamma_{\rm ev}$ at different times are shown in
Fig. 6.  
Effective adiabatic coeffcients, $\gamma_{\rm ev}$, are always nearly
equal to 4/3 in the case of quasi-static contraction.  
The reason for this is that the relation between $\gamma_{\rm ev}$ and 
$\gamma_{\rm{ad}}~(=7/5)$ is  
\begin{equation}
\label{eq:gamrel}
\gamma_{\rm ev}=\gamma_{\rm{ad}}-\frac{t_{\rm{qsc}}}{t_{\rm{cool}}},
\end{equation}
where $t_{\rm{cool}}=\varepsilon/(\partial L/ \partial m)$ and
$\varepsilon (\equiv \frac{1}{\gamma_{\rm{ad}}-1} 
\frac{kT}{\mu m_{\rm H}})$ is internal energy per unit mass. 
From the virial theorem,
\begin{equation}
3(\gamma_{\rm{ad}}-1)E_{\rm{i}}+E_{\rm g}=0,
\end{equation}
we find that 
\begin{equation}
\label{eq:tvir1}
\left|E_{\rm{i}}/\frac{dE_{\rm{i}}}{dt} \right|=\left|E_{\rm
    g}/\frac{dE_{\rm g}}{dt} \right| ,
\end{equation}
where $E_{\rm{i}}$ and $E_{\rm g}$ are the total internal energy and
gravitational energy:  
\begin{equation}
E_{\rm{i}}=\int _{0} ^{M} \varepsilon dm,   
~~~E_{\rm g}=- \int_{0}^{M} \frac{Gm}{r} dm .
\end{equation}
In the case of homologous contraction, 
$E_{\rm g} \propto \rho^{1/3}$, where we
compare $\rho$ for a fixed mass element (a homologous point), 
and $t_{\rm{qsc}}$ becomes constant along the radial coordinate: 
\begin{equation}
\label{eq:tvirial2}
t_{\rm{qsc}}=\left(\rho/\frac{d \rho}{dt} \right)=\frac{1}{3}
\left|E_{\rm g}/\frac{dE_{\rm g}}{dt} \right|. 
\end{equation}
On the other hand, by using total energy of a cloud,
\begin{equation}
  W=E_{\rm{i}}+E_{\rm g}=3\left( \frac{4}{3}-\gamma_{\rm{ad}}
  \right)E_{\rm{i}} , 
\end{equation}
its total luminosity can be given as
\begin{equation}
L=\frac{dW}{dt}=3\left(\gamma_{\rm{ad}}-\frac{4}{3} \right) \left|
  \frac{d E_{\rm{i}}}{dt} \right| .
\end{equation}
The mean cooling time-scale becomes  
\begin{equation}
\langle t_{\rm{cool}}
\rangle=\frac{E_{\rm{i}}}{L}=\frac{1}{3(\gamma_{\rm{ad}}-\dfrac{4}{3})} 
\left|E_{\rm{i}}/\frac{dE_{\rm{i}}}{dt} \right|. 
\end{equation}
Therefore, for homologously contracting spherical clouds, we have 
\begin{equation}
\label{eq:tsvir}
t_{\rm{qsc}}=\left(\gamma_{\rm{ad}}-\frac{4}{3} \right) \langle
t_{\rm{cool}} \rangle.  
\end{equation}
For the case of non-homogeneous contraction, this 
equation is not exactly valid. 
However, holds approximately in this case. 
From Eq.(\ref{eq:gamrel}), (\ref{eq:tsvir}) and identifying global
cooling time-scale $ \langle t_{\rm{cool}} \rangle$ with local one
$t_{\rm{cool}}$,  
$\gamma_{\rm ev}$ becomes nearly equal to $\frac{4}{3}$ as a whole. 
From Fig. 6, it can be seen that the initial steep gradients of the
$\gamma_{\rm ev}$ are levelled, and they assume values
around 4/3. 
This is because there exists a limiting structure for
clouds contracting quasi-statically, and then the contraction becomes
roughly homologous.
\section{Interpretation of results}
In \S 3 above, we have shown that the stellar mass molecular cores 
collapse dynamically, although some lines are optically thick at line
center. 
This fact indicates that the efficiency of cooling by H$_{2}$ lines does
not fall significantly even in such a situation, in contrast to the case 
of continuous radiation.   
The reason for this can be interpreted as follows.

Figure 7 displays the change of a processed profile of a single optically
thick line (optical depth at line center across the cloud is 76) along a ray
passing though the center of the cloud. 
Since the cloud is optically thick at line center, the specific
intensity at line center is saturated with the blackbody value
determined from the local temperature.  
On the other hand, optical depth at wing is smaller than unity. 
The final processed profile at the cloud surface is 
double peaked, and the peak frequencies correspond to 
frequencies where the optical depth across the core $\tau_{\nu,\rm{core}}>1$
(i.e., the radiation is saturated with the value of the core
temperature) and that of envelope $\tau_{\nu,\rm{env}}<1$ (i.e., the
radiation can be larger than the value of the envelope temperature). 
The height (i.e., the normalization of intensity) and width of these
peaks are of the same order of magnitude as the blackbody value
determined from the temperature near the center, not around the surface
of the cloud.  
The reduction factors from the central values are only several. 
Owing to the effect described above, which is a distinctive feature of
line cooling, the temperature around the cloud center 
can be `seen' from outside the cloud, even when lines are optically
thick at line center and the surface temperature is relatively
low. Accordingly, its cooling proceeds  
efficiently, as long as there are enough frequency ranges 
where $\tau_{\nu,\rm{core}}>1$ and $\tau_{\nu,\rm{env}}<1$.
Therefore, the decrease in the efficiency of cooling 
caused by the low surface temperature is not severe in the line
cooling case, in contrast to the case of cooling by continuum.

Figure 8 displays the ratio of $L_{i}$, 
the contribution to the luminosity at the
surface from each line $i$, to $L_{{\rm thin},i}$, 
the value in the ``optically thin'' case 
(i.e., $L(r_{s})=\sum_{i={\rm all~lines}} L_{i}$, and
     $L_{{\rm thin},i}=4 \pi \int_{{\rm whole~cloud}} dV \int_{{\rm line}~i}
     d\nu j_{\nu}$),  
versus optical depths at
line center across the clouds $\tau_{\rm{c,tot}}$ for each line. 
All marks of a given variety represent lines of a single given cloud,
and the solid curve represents the value of $L_{i}/L_{{\rm thin},i}$ for
the uniform cloud of $1M_{\odot}$, whose number density equals 
$10^{11}~{\rm cm^{-3}}$.      
From this figure, it can be seen that 
$L_{i} / L_{{\rm thin},i}$ falls only as $\sim {\tau_{\rm{tot}}}^{-1}$. 
This simply reflects the fact that in the ``optically thin'' case, the 
specific intensity along a ray at optical depth $\tau$ (even if $\tau >
1$) is determined as 
\begin{equation}
I_{\nu,\rm{thin}}=\int_{0}^{\tau} S_{\nu}(\tau')d\tau',
\end{equation}
where $S_{\nu}=j_{\nu}/\alpha_{\nu}$ is the source function.
On the other hand, in the optically thick case,
\begin{equation}
I_{\nu}\sim S_{\nu}(\tau).
\end{equation}
Then, if $\tau > 1$
\begin{equation}
I_{\nu}/I_{\nu,\rm{thin}} \sim \tau^{-1}.
\end{equation}
Therefore, no significant suppression of cooling caused by the low surface
temperature arises in Fig. 8. 

As discussed above, the effects of the surface temperature are not severe even
for optically thick lines. Thus it is appropriate to use the
value of a homogeneous cloud in estimating the cooling time. This
is given by
\begin{equation}
t_{\rm{cool}}=\frac{E_{\rm{i}}}{L}=\frac{M c_{\rm v} T}{4 \pi R^{2} \sigma T^{4} f}
\propto M^{1/3} T^{-3} n^{2/3} f^{-1}, 
\end{equation}   
where $E_{\rm{i}},M,T,R,c_{\rm v}$ and $n$ are the internal energy, the
mass, the 
temperature, the radius, the specific heat at constant volume, and the
number density of the homogeneous cloud, 
respectively. The quantity $f$ is the ratio of the luminosity of the cloud to
the blackbody luminosity of the temperature $T$. 
In the case that cooling is dominated by line cooling, by 
using the line width $\Delta \nu$ and the effective number
of lines $\alpha_{\rm{c}}$, $f$ can be written as 
$f=L/4 \pi R^{2}\sigma T^4 \sim \alpha_{\rm{c}} 
\frac{\Delta \nu}{\nu}$. 
We adopt the Doppler width $\frac{\Delta 
  {\nu}_{\rm{D}}}{\nu }=\frac{1}{c} \sqrt{\frac{2kT}{\mu m_{\rm H}}}$ 
for the line width. 
Assuming the cloud is almost in hydrostatic equilibrium initially, we 
can relate the temperature with the mass and the number density:
$T \propto M^{2/3} n^{1/3}$. From these relations, we obtain
\begin{equation}
t_{\rm{cool}}\propto M^{-2} n^{-1/2} \alpha_{\rm{c}}^{-1}.
\end{equation}
On the other hand, the free-fall time is given by
\begin{equation}
t_{\rm{ff}}=\sqrt{\frac{3 \pi}{32 G \rho}} \propto n^{-1/2}.
\end{equation}
Then the ratio of these two time-scales is
\begin{equation}
\frac{t_{\rm{ff}}}{t_{\rm{cool}}} \propto M^{2} \alpha_{\rm{c}}.
\end{equation}
As the cloud contracts, the number of optically thick lines (in this
case, $\alpha_{\rm{c}}$) increases, and the ratio
$t_{\rm{ff}}/t_{\rm{cool}}$ becomes 
larger. This accounts for the results of numerical calculation described
above (see also Ref.~\citen{rf:4}, and \citen{rf:15}).  
\section{Conclusion and Discussion}
Our principal result is that if cooling is dominated by 
$\rm {H_2}$ line cooling, a primordial cloud of stellar mass scale is
able to cool efficiently.  
This is because the efficiency of line cooling does not decrease
significantly by the effect of a line profile, even when the cloud is
optically thick for line center frequency and its surface temperature is relatively low. 
In particular, fragments of filamentary clouds collapse dynamically
after fragmentation. 
Although our calculation is limited to
$\sim 1 M_{\odot}$ composed only of ${\rm H_2}$, 
our results can be applied to more general 
molecular cores, as long as cooling is dominated by optically thick 
line emissions.

A dynamically collapsing spherical gas cloud is unstable to a
non-spherical perturbation and becomes disk-like.\cite{rf:16} 
Therefore, dynamically collapsing fragments may also become disk-like.  
Since a self-gravitating disk is unstable to fragmentation, 
cloud cores may fragment again. 
Whether repeated fragmentations are possible or not
depends on the thickness of the disk. 
Fragmentation occurs for a thin disk
(axis ratio $>2\pi$), but does not occur for a thick disk.\cite{rf:17} 
Because fragments of
filamentary clouds start collapsing from nearly hydrostatic equilibrium, 
it seems unlikely that they reach a highly flattened configuration. 
Therefore it is expected that even if fragmentation were to occur,
it would not be the case that numerous small pieces would be produced.
  
Uehara et al.\cite{rf:4} investigated the fragmentation of filamentary
clouds with a one-zone approximation. 
Recently, Nakamura and Umemura\cite{rf:18} examined 
cylindrical collapse of primordial clouds with radial structures and
showed that at the epoch of fragmentation the clouds display clear
core-envelope structures. 
They suggested that Uehara et al.'s\cite{rf:4} one-zone clouds
correspond to the cores of their clouds. 
Hence, if the contraction of fragments were slow, the cores would
accrete ambient gas. 
This would result in more massive ones. 
However, our analysis shows the contraction is sufficiently rapid
(its time-scale is the free-fall time-scale in the core). 
Thus the accretion of the envelope (its time-scale is the free-fall
time-scale in the envelope), if it exists, becomes 
significant only after the molecular core contracts sufficiently into a
stellar core. The accretion of the envelope does not affect the dynamics of
the contracting core significantly. Consequently, it is possible to treat the
contraction of the core and the accretion of ambient gas separately.

In this work, we showed that fragments of filamentary clouds collapse
dynamically, although we have not pursued yet the actual evolution of the
cloud after they begin dynamical collapse. 
To elucidate the actual contraction of primordial protostellar clouds,
elaborate hydrodynamical calculations are needed. 
This will be presented in a forthcoming paper.\cite{rf:19}   

\section*{Acknowledgements}
We would like to thank Professor H. Sato for continuous
 encouragement and Professor N. Sugiyama for
 critical reading of the manuscript. 
This work is supported in part by Research Fellowships of the Japan Society 
for the Promotion of Science for Young Scientists, No.6894~(HU), 2370~(HS)
 and by the Grant-in-Aid for Scientific Research from the Ministry of
 Education, Science, Sports and Culture No.~09740174~(RN).



\newpage
\begin{figure}
  \caption{Ray geometry of radiative transfer calculation in spherical
  symmetry. We solve frequency-by-frequency radiative transfer equations
  along rays tangent to each mass shell and obtain the intensity
  $I_{\nu}(\theta)$ at each mesh point. }
\end{figure}

\begin{figure}
\caption{The contribution from each line to the luminosity at the
surface versus its wavelength.
The mass of the cloud is $1 M_{\odot}$, the number density at half mass
radius is $10^{11}~{\rm cm}^{-3}$, and the density/temperature
distribution is represented by the Emden function of polytropic index
2.5.}
\end{figure}

\begin{figure}
\caption{Contours of the ratio $t_{\rm{ff}}/t_{\rm{qsc}}$ at the center of
  clouds of polytropic index (a) $N=2.5$ and (b) $N=8$. 
  The contour spacing is logarithmic with an increment of 0.5. 
  The upper-right regions correspond to the temperature $T>2000~{\rm
  K}$, where hydrogen molecules begin to dissociate.}
\end{figure}

\begin{figure}
\caption{Evolution of the effective exponent for structure,
  $\gamma_{\rm str}(m) \equiv \frac{(\partial {\rm ln} p/ \partial 
  m)_{t}}{(\partial {\rm ln} \rho / \partial m)_{t}}$, for two
  quasi-statically contracting clouds. 
  The initial conditions (solid lines) are
  (a)~$M=0.5M_{\odot},~n_{h}=10^{9}~{\rm
  cm^{-3}},~N=2.5$. (b)~$M=0.2M_{\odot},~n_{h}=10^{9}~{\rm cm^{-3}},~N=20$. 
  Each curve depicts $\gamma_{\rm str}(m)$ at some specific time (in order of
  time: solid, dotted, short-dashed, long-dashed and dash-dotted line).
  The central temperature and density are (a) initial: $149~{\rm K}, 1.35
  \times 10^{-14}~{\rm g cm^{-3}}$, final: $2048~{\rm K}, 1.50 \times
  10^{-10}~{\rm g cm^{-3}}$ and (b) initial: $89~{\rm K}, 2.48 \times
  10^{-14}~{\rm g cm^{-3}}$, final: $2154~{\rm K}, 4.41 \times
  10^{-9}~{\rm g cm^{-3}}$.  
  These values approach $1.12-1.13$ around the central region for both
  clouds when the central temperature reaches $\sim 2000$~K. 
  Note that both
  clouds start to dynamical collapse before reaching the final states shown
  in the figure. Therefore, our quasi-static formulation cannot be
  applied, and thereafter the evolution shown in the figure does not
  correspond to the actual evolution of the clouds. }
\end{figure}

\begin{figure}
\caption{Evolution of free-fall time $t_{\rm{ff}}$ and dynamical time for
  quasi-static contraction $t_{\rm{qsc}}$ at the center and their ratio
  for a cloud which contracts quasi-statically at the beginning
  (case (a) of Fig.4). The time-scales $t_{\rm{ff}}$ and
  $t_{\rm{qsc}}$ and the time ellapsed from the begining $t$ are all
  normalized by $t_{\rm{qsc},0}$, the initial value of $t_{\rm{qsc}}$.
  This figure indeed shows the cloud begins to
  collapse dynamically after $\sim 1.5 t_{\rm{qsc},0}$.}
\end{figure}

\begin{figure}
\caption{Evolution of the effective exponent for temporal evolution,
  $\gamma_{\rm ev}(m) \equiv \frac{(\partial {\rm ln} p/ \partial 
  t)_{m}}{(\partial {\rm ln} \rho / \partial t)_{m}}$, for the same
  clouds as Fig.4. Here the same line corresponds to the same time as
  Fig.4.}
\end{figure}

\begin{figure}
\caption{Processed line profiles of an optically thick line along a
  ray passing through the center of the cloud at the
  radii containing 0 \% (center, solid line), 1.4 \% (dotted line),
9.9 \% (short dashed line), and 100 \% (surface, long dashed line) of
  the total mass. 
The optical depth at the line-center frequency $\tau_{\rm{c}}=38, 46, 54
  \rm{and} 76$, respectively,
  measured from the surface of incidence. 
 The line corresponds to the transition from $(v=1,~j=3)$ to
  $(v=0,~j=1)$. 
  The parameters of the cloud are the same as in Fig.2. 
  The abscissa represents frequency from line center 
  normalized by the Doppler width at the center of the cloud $\Delta
  \nu_{\rm{D,c}}$.}
\end{figure}

\begin{figure}
\caption{The ratio of $L_{i}$, the contribution to the luminosity at the
surface from each line $i$, to $L_{{\rm thin},i}$, the value in the
``optically thin'' 
case, as a function of the optical depth at
line-center frequency across the clouds $\tau_{\rm{c,tot}}$ for each
line. 
These relations are demonstrated for two clouds. Open circles and crosses
correspond to lines 
from the cloud of $(N=2.5,~M=1M_{\odot},~n_{h}=10^{11}~{\rm cm}^{-3})$
and $(N=8,~M=1M_{\odot},~n_{h}=10^{11}~{\rm cm}^{-3})$, respectively. 
The solid curve represents $L_{i}/L_{{\rm thin},i}$ for uniform clouds.}
\end{figure}

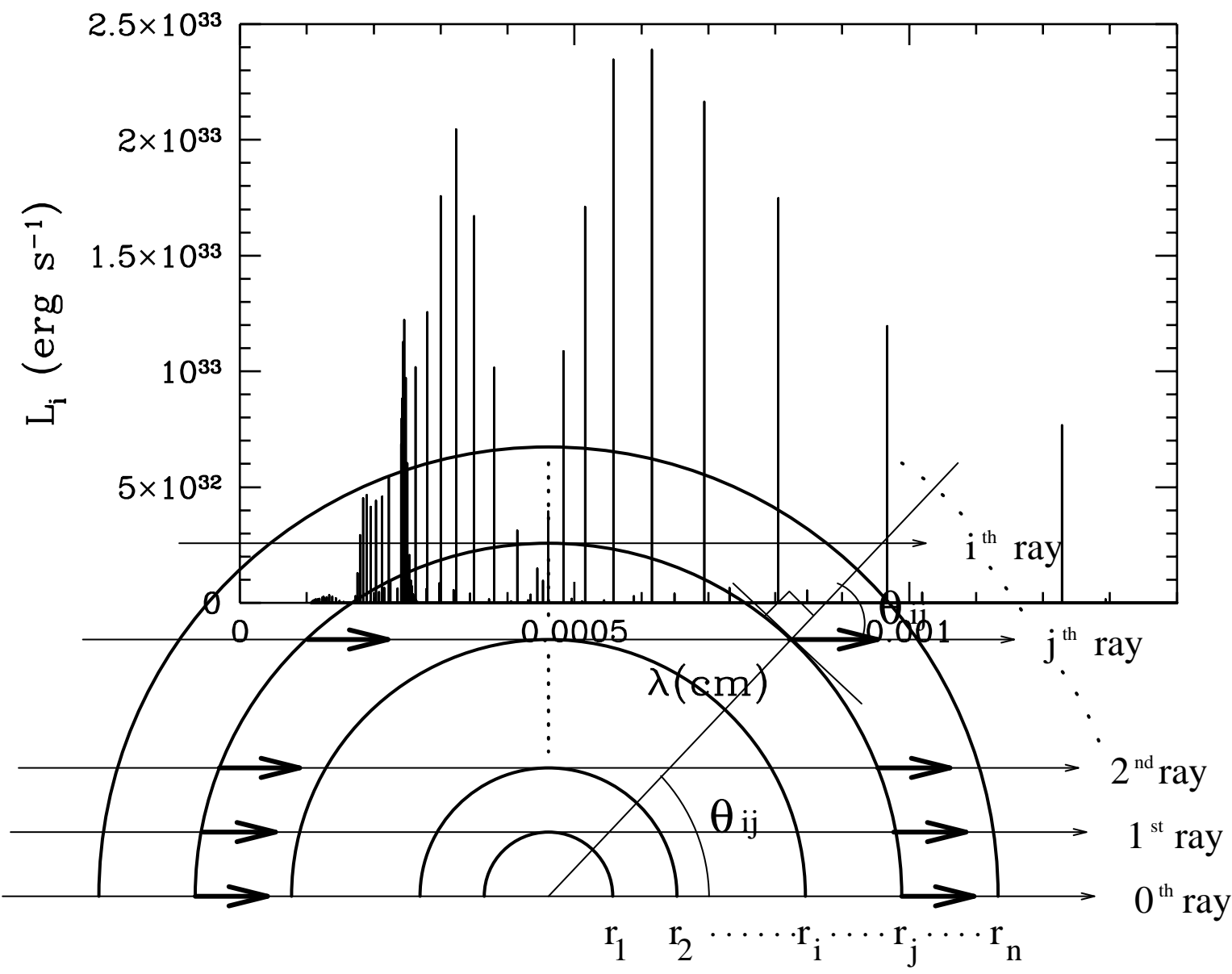
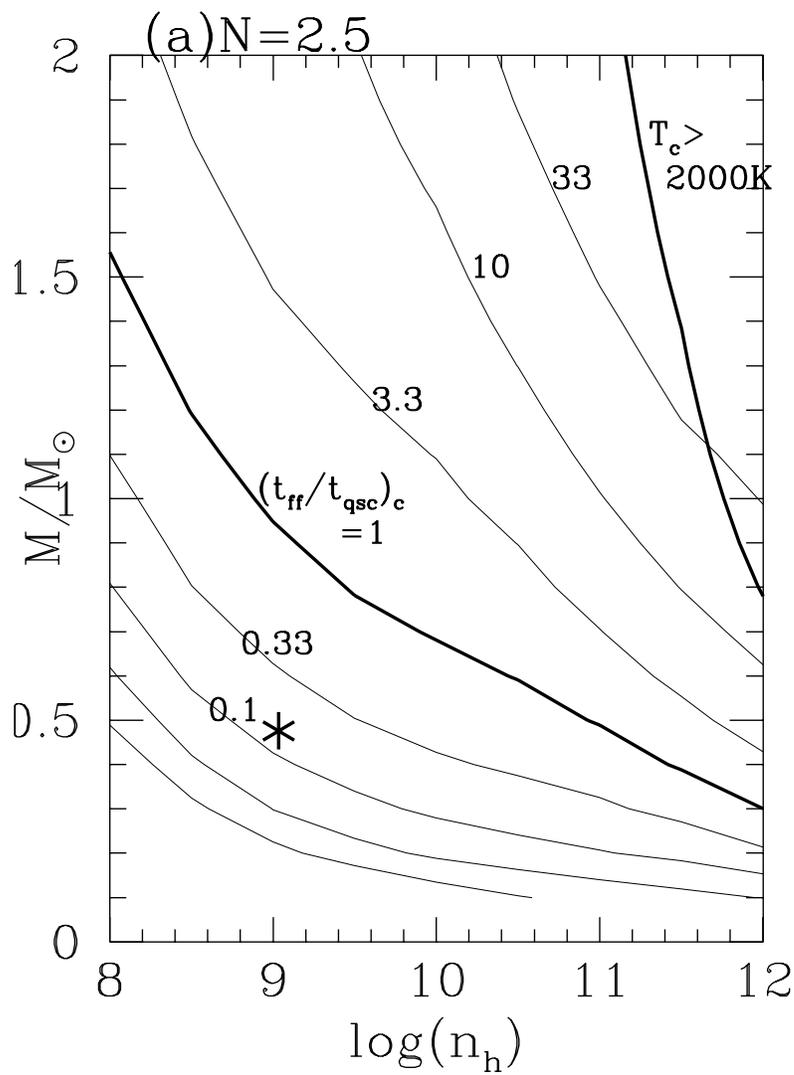
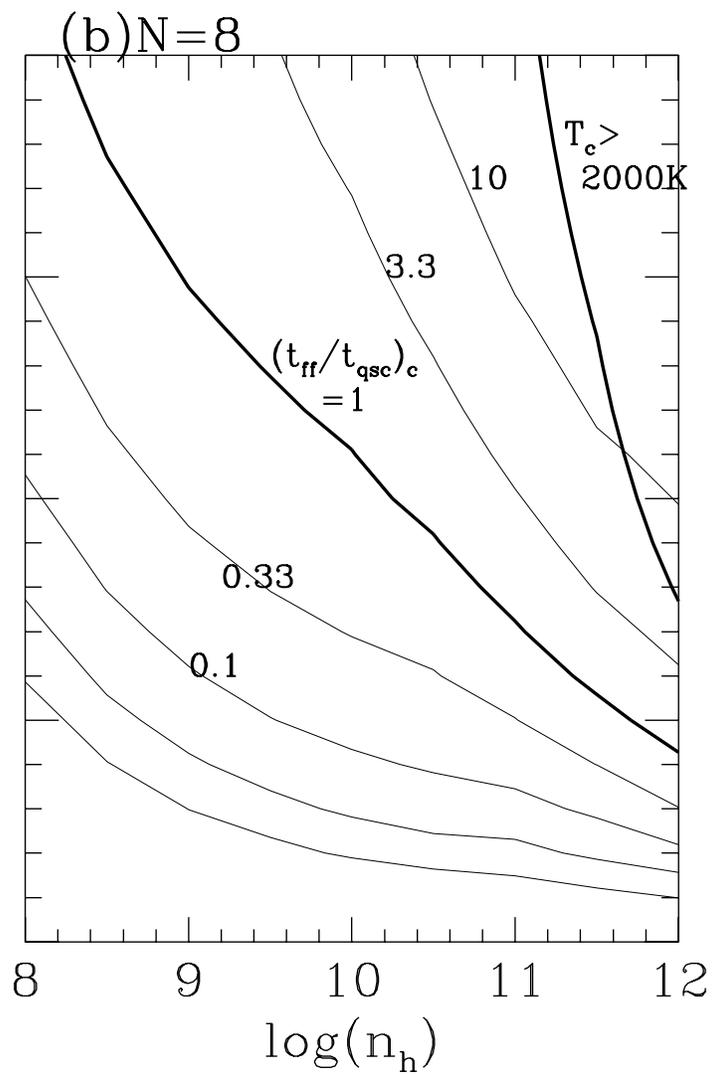
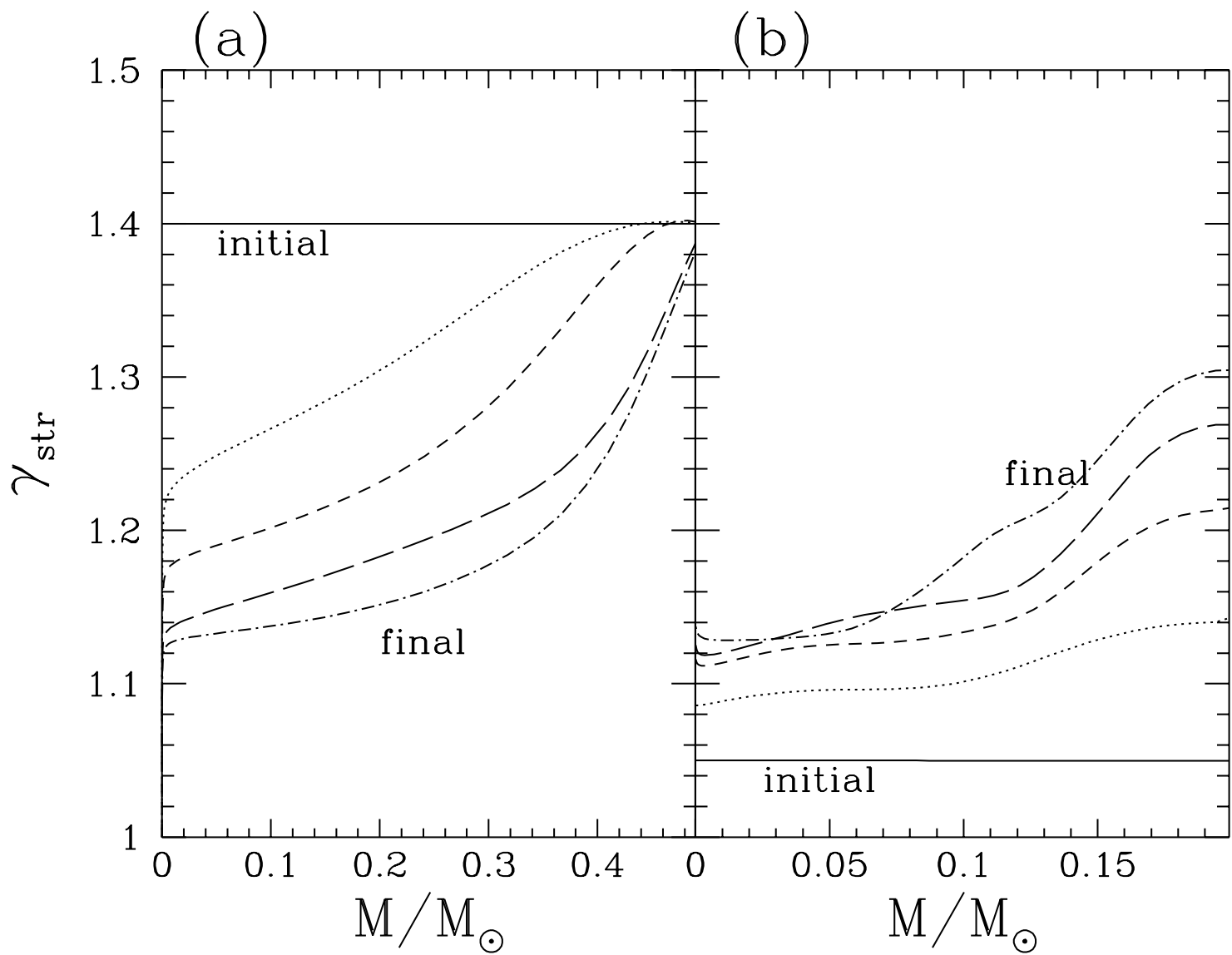
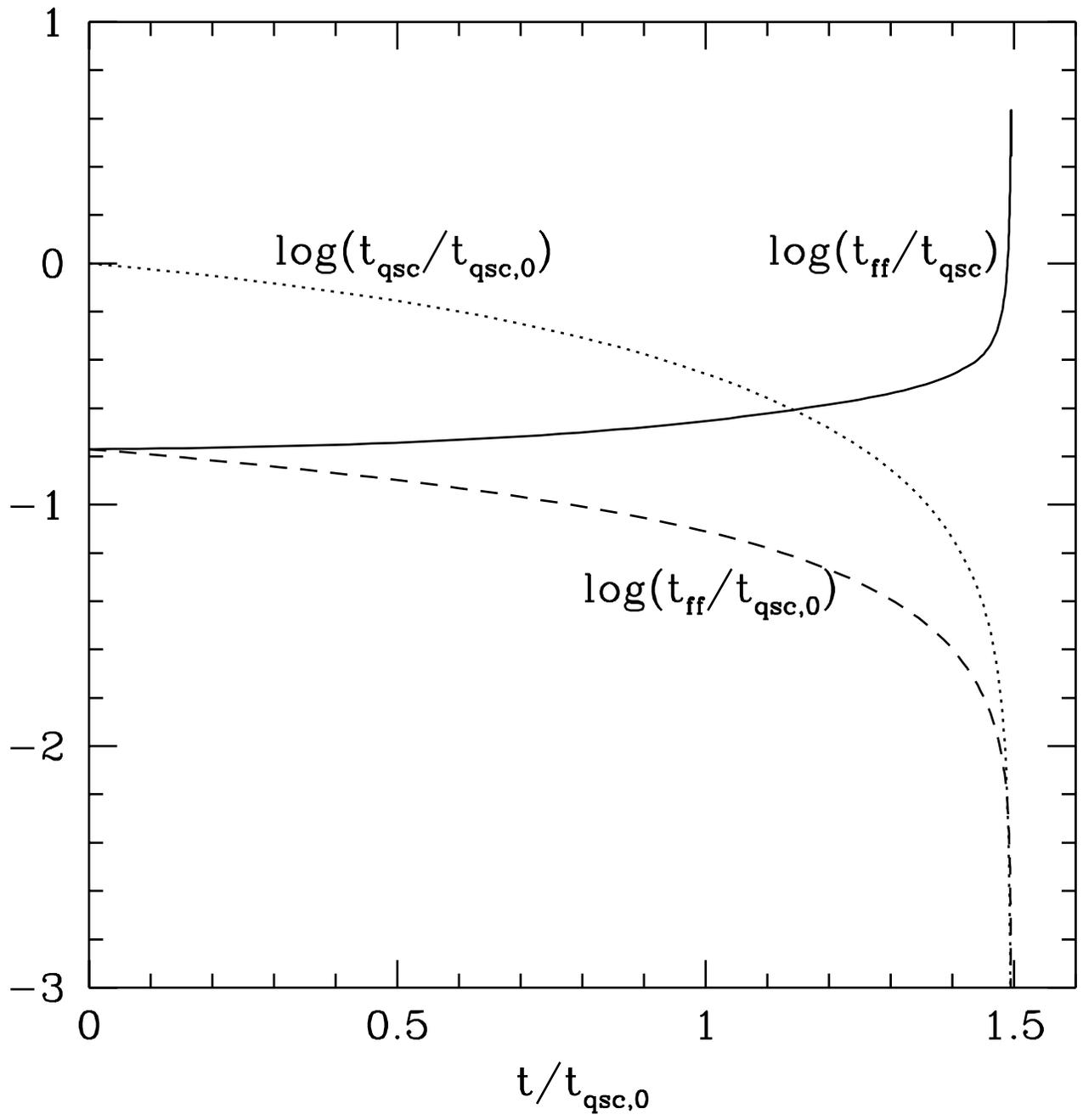
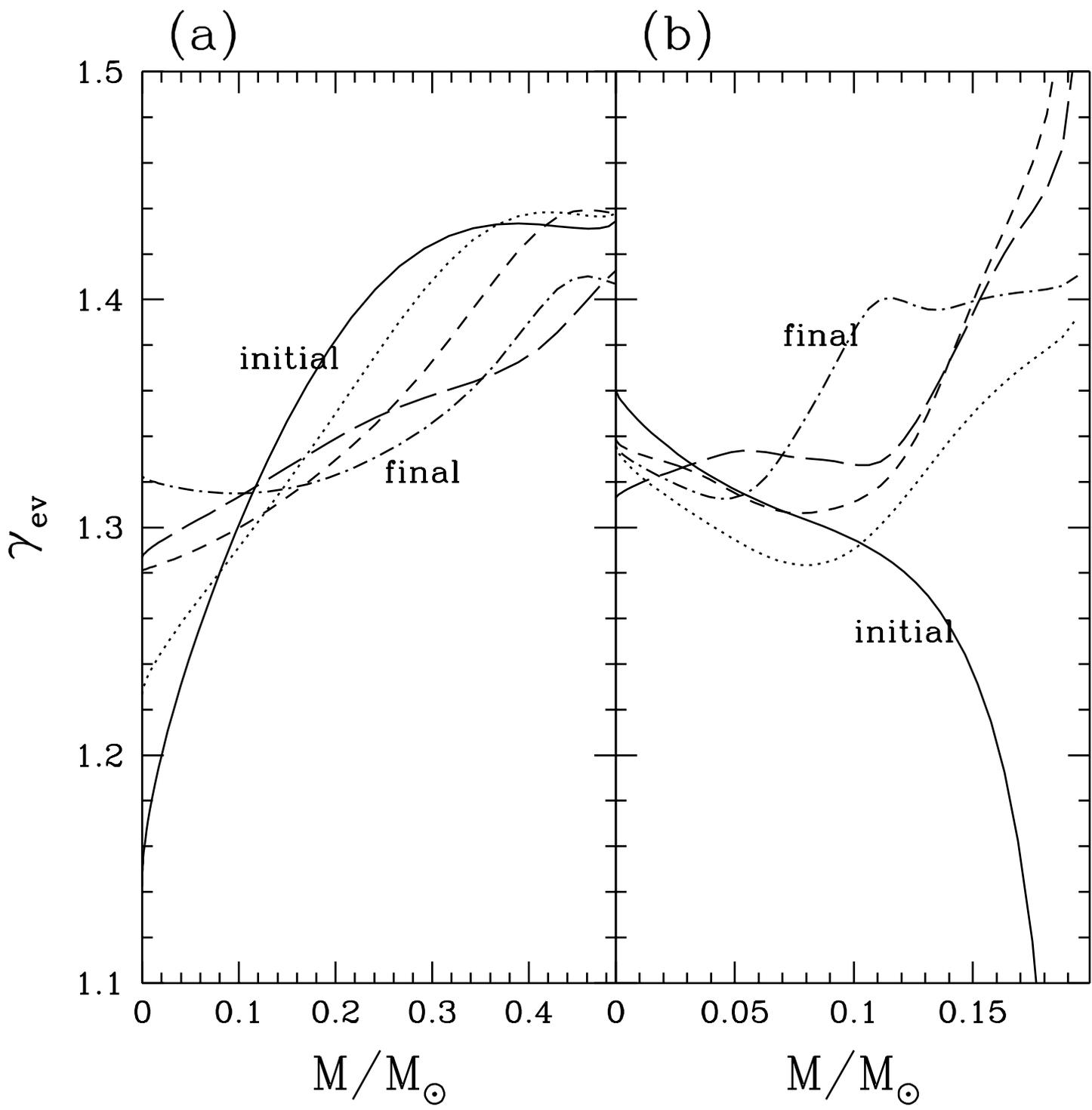
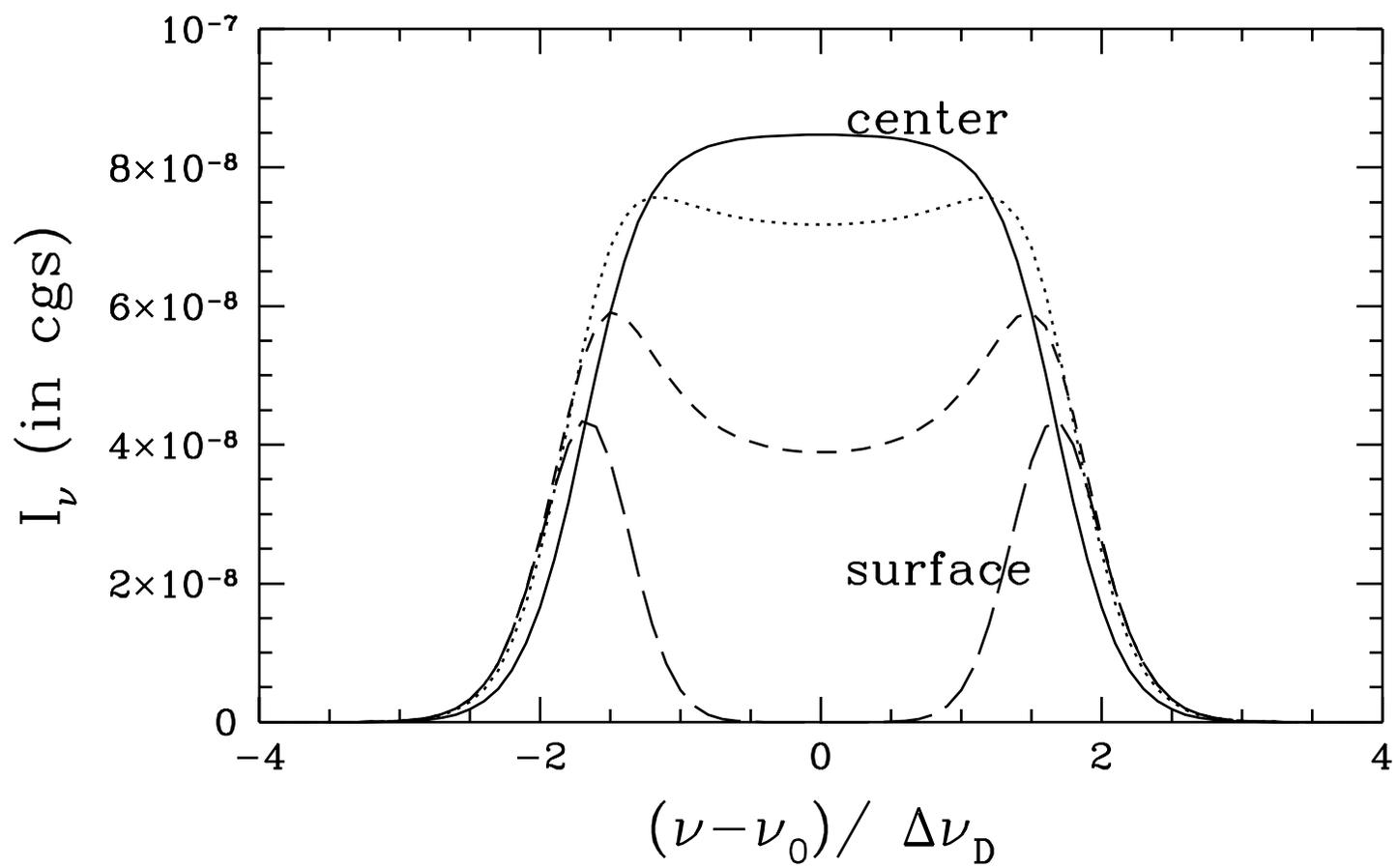
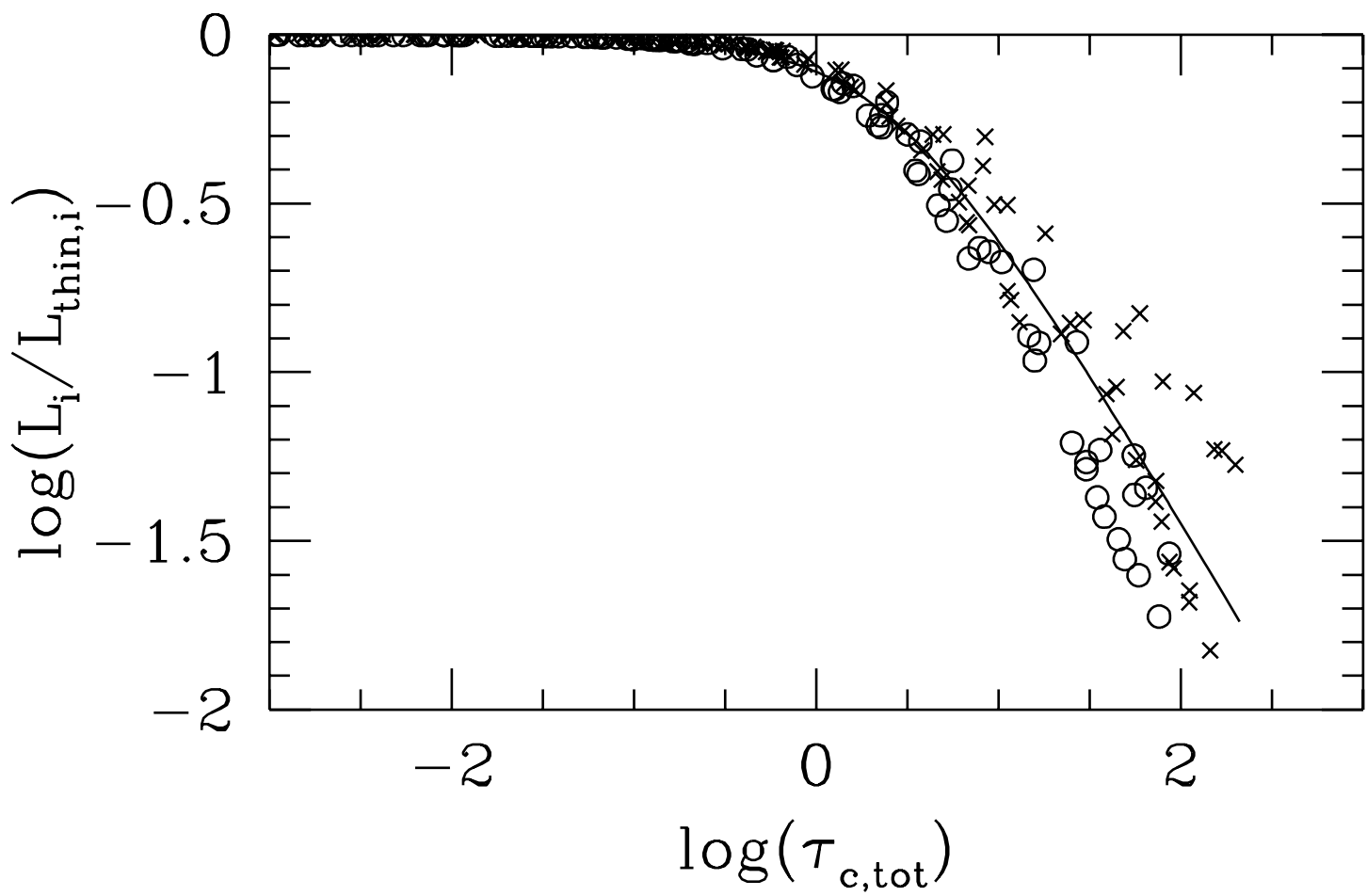
\end{document}